\documentstyle[12pt,epsf]{article}
\newcommand{\He}{^3\mbox{He}}

\newcommand{\ben}{\begin{displaymath}}
\newcommand{\een}{\end{displaymath}}
\newcommand{\be}{\begin{equation}}
\newcommand{\ee}{\end{equation}}
\newcommand{\bea}{\begin{eqnarray}}
\newcommand{\eea}{\end{eqnarray}}

\newcommand{\piNN}{$\pi N\!N$}

\newcommand{\NNN}{$N\!N\!N$}

\newcommand{\tZ}{\tilde{Z}}

\newcommand{\bc}{\begin{center}}
\newcommand{\ec}{\end{center}}

\newcommand{\eqn}[1]{\label{#1}}
\newcommand{\eq}[1]{Eq.\ (\ref{#1})}

\newcommand{\fign}[1]{\label{#1}}

\begin{document}
{\Large\bf Gauging the three-nucleon system }\footnote{Contributed talk at XVth
International Conference on Few-Body Problems in Physics, Groningen, 1997,
http://www.kvi.nl/disk\$1/fbxv/www/abs\_list\_num.html} \\[2mm]
{\large\em A.N.\ Kvinikhidze,$^\dagger$ B.\ Blankleider}\\[2mm]
{\small Physics Department, Flinders University of South Australia,
Bedford Park, S.A. 5042, Australia}\\[2mm]
{\bf Introduction}\\
We show how to attach an external photon to a system of three strongly
interacting nucleons described by three-body scattering equations. Our method
involves the idea of gauging the scattering equations themselves, and results in
electromagnetic amplitudes where the external photon is effectively coupled to
every part of every strong interaction diagram in the model. Current
conservation is therefore implemented in the theoretically correct fashion. In
this way we obtain the expressions needed to calculate all possible
electromagnetic processes of the three-nucleon system including the
electromagnetic form factors of the three-body bound state, $pd\rightarrow
pd\gamma$, $\gamma ^3\mbox{He}\rightarrow pd$, $\gamma \He\rightarrow ppn$,
etc. As the photon is coupled everywhere in the strong interaction model, a
unified description of the \NNN -$\gamma$\NNN\ system is obtained. An 
interesting aspect of our results is the natural appearance of subtraction terms
needed to avoid the overcounting of diagrams. We have also applied the gauging
of equations method to the four-dimensional description of the \piNN\ system
which, because of the possibility of pion absorption, has overcounting problems
all of its own, and is therefore much more complicated than the \NNN\ case
discussed here. Despite these complications, the gauging procedure itself is
very simple and leads easily to a unified description of the $\pi N\!N$-$\gamma
\pi N\!N$ system.$^1$
\medskip

{\bf Gauging the three-nucleon Green function}\\
In this presentation we work within the covariant framework of quantum
field theory. For three distinguishable nucleons, the Green function $G$ is
expressed in terms of its fully disconnected part $G_0$, and the kernel $V$, by
\be
G=G_0+G_0VG . \eqn{G}
\ee
This equation is basically a topological statement regarding the three-particle
irreducible structure of Feynman diagrams belonging to $G$; as such, it can be
utilised directly to express the structure of the same Feynman diagrams, but
with photons attached everywhere. Thus from \eq{G} it immediately follows that
\be
G^\mu =G_0^\mu + G_0^\mu VG+G_0V^\mu G+G_0VG^\mu .  \eqn{G^mu}
\ee
This result expresses $G^\mu$ in terms of an integral equation, and illustrates
what we mean by {\em gauging an equation}, in this case the gauging of \eq{G}.
Note that $G^\mu$ consists of all the Feynman diagrams belonging to $G$ where a
photon has been attached to all possible places. Neglecting three-body forces,
we follow the usual formulation of the three-body problem and write
$V=V_1+V_2+V_3$ with $V_i = v_i d_i^{-1}$ where $d_i$ is the propagator of
nucleon $i$, and $v_i$ is the two-body potential between particles $j$ and $k$,
($i\ne j,k$). By solving \eq{G^mu} for $G^\mu$ and further gauging the above
equation for $V_i$, we obtain that
\be
G^\mu = G\Gamma^\mu G
\ee
where $\Gamma^\mu$ is the three-nucleon electromagnetic vertex function given by
\be
\Gamma^\mu=\sum_{i=1}^3 \left(\Gamma_i^\mu D_{0i}^{-1}+v_i^\mu d_i^{-1}
-v_i\Gamma_i^\mu \right)  .\eqn{hello}
\ee
Here $D_{0i}$ is the free two-body Green function for particles $j$ and $k$,
$\Gamma_i^\mu$ is the $i$'th nucleon's electromagnetic vertex function, and
$v_i^\mu$ is the gauged two-body potential. \eq{hello} is illustrated in Fig.~1.
\begin{figure}[t]
\hspace*{1cm}  \epsfxsize=13cm\epsfbox{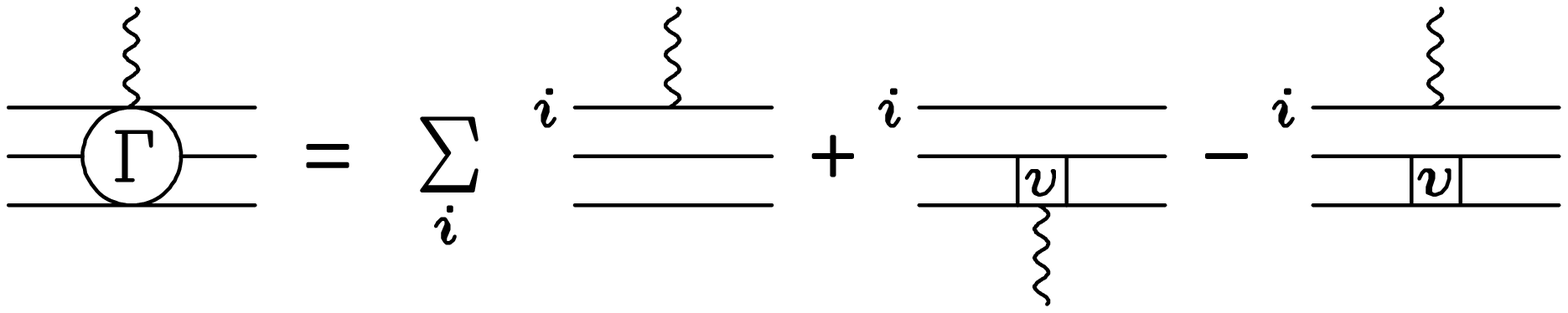}
\caption{\fign{gamma_3d} Illustration of \protect\eq{hello} for the
three-nucleon electromagnetic vertex function $\Gamma^\mu$.}
\end{figure}
It is interesting to note the natural appearance of the subtraction term
$-v_i\Gamma_i^\mu$. When used to calculate $G^\mu$, the presence of this
subtraction term in $\Gamma^\mu$ stops the overcounting of diagrams coming from
the relativistic impulse approximation term (first term on the r.h.s. of
\eq{hello}). By taking appropriate residues of Eq.~3, we can obtain expressions
for all possible electromagnetic observables of three nucleons. In the
pioneering four-dimensional calculations of the \NNN\ system by Rupp and
Tjon$^2$, the electromagnetic form factors were calculated by analogy with
three-dimensional descriptions where there is no subtraction term; as a result,
these calculations contain overcounting.
\medskip

{\bf Gauging the AGS equations}\\
Although Eqs.\ (3) and (4) solve the problem of gauging three nucleons, they do
require explicit knowledge of the two-body potential $v_i$ and the gauged
two-body potential $v_i^\mu$. As it is often preferable to have the input in
terms of the two-body t-matrix $t_i$ and the gauged two-body t-matrix $t_i^\mu$,
we provide an alternative solution to the gauging problem.  Our starting point
here shall be the AGS equations describing the scattering of three identical
nucleons. The essential point is that the input to these equations consists of
two-body t-matrices. For identical particles there is a variety of ways to
define the AGS amplitude, all giving the same three-body Green function $G$.
The one we have chosen, which we call $Z$, satisfies the particularly simple 
AGS equation for identical particles
\be
\tZ = G_0 + D_{03}t_3{\cal P}\tZ   \eqn{Z}
\ee
where $\tZ=G_0ZG_0$ and where ${\cal P}$ shifts particle labels cyclically to
the left.  As $\tZ$ is related to the three-body Green function $G$ in a simple
way, the solution to our problem of specifying $G^\mu$ now rests essentially on
the construction of the gauged AGS Green function $\tZ^\mu$. Gauging \eq{Z} in
the same way we gauged \eq{G}, we obtain \be \tZ^\mu=\tZ d_3^{-1}d_3^\mu
+\tZ\left( D_{03}^{-1}D_{03}^\mu D_{03}^{-1}d_3^{-1} +d_3^{-1}t_3^\mu{\cal
P}\right) \tZ .  \eqn{Zmu} \ee This equation describes the attachment of photons
at all possible places in the multiple-scattering series of three identical
particles. The input is in terms of two-body t-matrices. As such, it forms the
central result in the gauged three-nucleon problem.
\vspace{1cm}

{\small
$^\dagger$ On leave from Mathematical Institute of Georgian Academy of
Sciences, Tbilisi, Georgia.\\
1.\ Kvinikhidze AN, Blankleider B, {\em Coupling photons to
hadronic processes}, invited talk at the Joint Japan Australia Workshop,
Quarks, Hadrons and Nuclei, November 15-24, 1995 (unpublished); a more detailed
account is in preparation.\\
2.\ Rupp G, Tjon JA, Phys.\ Rev.\ C {\bf 37} (1988) 1729; {\em ibid.} {\bf 45}
(1992) 2133.
}
\end{document}